\documentclass[twocolumn,aps]{revtex4}
\usepackage{amsmath}
\usepackage{graphicx}

\begin{document}

\title{Asymmetric radiating brane-world}

\author{L\'{a}szl\'{o} \'{A}. Gergely$^{1,2}$, Emily Leeper$^1$,
Roy Maartens$^1$}

\affiliation{$^1$Institute of Cosmology and Gravitation,
University of Portsmouth, Portsmouth~PO1~2EG, UK\\
$^2$Departments of Theoretical and Experimental Physics,
University of Szeged, Szeged~6720, Hungary}

\begin{abstract}

At high energies on a cosmological brane of Randall-Sundrum type,
particle interactions can produce gravitons that are emitted into
the bulk and that can feed a bulk black hole. We generalize
previous investigations of such radiating brane-worlds by allowing
for a breaking of $Z_2$-symmetry, via different bulk cosmological
constants and different initial black hole masses on either side
of the brane. One of the notable features of asymmetry is a
suppression of the asymptotic level of dark radiation, which means
that nucleosynthesis constraints are easier to satisfy. There are
also models where the radiation escapes to infinity on one or both
sides, rather than falling into a black hole, but these models can
have negative energy density on the brane.

\end{abstract}

\date{\today}

\maketitle

\section{Introduction}

Brane-world models of the universe allow us to explore the
cosmological implications of extra-dimensional gravity. In these
models, matter is confined to a four-dimensional ``brane" while
the gravitational field propagates in a higher-dimensional ``bulk"
spacetime. Of particular interest are the Randall-Sundrum (RS)
type cosmological models~\cite{rev}, with one extra spatial
dimension that is infinite. At low energies the 5D graviton is
localized at the brane, because of the strong curvature of the
bulk due to a negative bulk cosmological constant, and general
relativity is recovered with small corrections. At high energies,
gravity behaves as a 5D field and there are large deviations from
general relativity. The effective 4D Planck scale on the brane,
$M_4\approx 10^{19}\,$GeV, can be much higher than the true 5D
Planck scale $M_5$ ($10^8\,\mbox{GeV}<M_5<M_4$).

The background RS-type cosmology is a Schwarzschild-anti de Sitter
(AdS) bulk with a Friedman-Robertson-Walker brane that is moving
away from the black hole as the universe expands. The mass
parameter $m$ of the bulk black hole produces a Coulomb effect at
the brane. This Coulomb effect is felt on the brane as a ``dark
radiation" term $m/a^4$ in the modified Friedman equation on the
brane. One of the corrections to the background model that can be
considered, is to incorporate the production of 5D gravitons in
particle interactions~\cite{hr}. This can become significant at
very high energies. The gravitons are emitted into the bulk and
feed the black hole, so that $\dot m>0$. At low energies the
effect is negligible and $m$ asymptotes to a constant value. The
general problem is highly complicated, but a simplified model can
be constructed by neglecting the back-reaction of the gravitons on
the bulk geometry, and by assuming that all gravitons are emitted
radially~\cite{LSR} (some progress in the case with non-radial
emission has been made~\cite{LS}). This leads to a Vaidya-AdS bulk
(previously considered as a model for thermodynamic interactions
between the brane and the black hole~\cite{ckn}). The dynamical
equations for the simple radiating brane-world may be solved
exactly~\cite{LMS}. The black hole mass grows monotonically and
tends to a late-time constant. An expanding brane is always
outside the black hole horizon~\cite{ms}.

Here we consider a further generalization of the radiating
brane-world model, by allowing for asymmetry in the bulk, i.e.,
different cosmological constants and initial black hole masses on
either side of the brane. Such asymmetry has been previously
investigated for non-radiating
brane-worlds~\cite{lamasym2,both1,lamasym1,perk,bhasym,both2}.
(The equations for the completely general case are derived
in~\cite{Decomp}.) We find that for radiating brane-worlds, the
introduction of asymmetry leads to some interesting new features,
in addition to those already identified previously for
non-radiating models.

\section{The bulk and brane}

The bulk metric is Vaidya-anti de Sitter~\cite{ckn},
\begin{eqnarray}
d{s}^{2} =-f( v,r) dv^{2}+2\epsilon\, dvdr
+r^{2}d\vec{x}^{\,2}\,, \label{ChVAdS5}
\end{eqnarray}
where $v=\,$const are plane-fronted null rays which are ingoing
for $\epsilon =1$ and outgoing for $\epsilon =-1$, and
\begin{equation}
f( v,r) =-\frac{2m( v) }{r^{2}}-\frac{{\Lambda }}{6}r^{2}\,.
\label{ff}
\end{equation}
Here $\Lambda$ is the bulk cosmological constant, and $m$ is the
mass parameter of the black hole at $r=0$. We assume that the
brane is outside the horizon ($f>0$), which requires negative
$\Lambda$ for non-negative $m$. The metric is a solution of the
bulk field equations
\begin{equation}
^{(5)\!}G_{ab}+\Lambda\,^{(5)\!}g_{ab}=\kappa_5^{2}
\,^{(5)\!}T_{ab}\,,
\end{equation}
where $\kappa_5^{2}=8\pi/M_5^3$, and the bulk stress tensor has
null-radiation form,
\begin{equation}
^{(5)\!}T_{ab}=\psi ( v,r) k_{a}k_{b}\,. \label{TND}
\end{equation}
Here $\psi$ gives the power density in the null radiation that
flows along the null rays generated by $k_a$, where $k_a \propto
\delta_a{}^v$. The field equations require that
\begin{equation}
\kappa_5^{2}\psi =\epsilon
\frac{3\dot{v}^{2}}{r^{3}}\,\frac{dm}{dv}\,, \label{psi}
\end{equation}
where $\dot v=dv/dt$ and $t$ is cosmic proper time on the brane.
The brane trajectory is $r=a(t)$, and we have taken the brane to
be spatially flat. The ray vector has been normalized so that
$k_au^a=-1$, where $u^a$ is the brane 4-velocity. From
$u^{a}u_{a}=-1$ we obtain
\begin{equation}
f\dot{v}=\epsilon \dot{r}+\left( \dot{r}^{2}+f\right) ^{1/2}\ .
\label{vdot}
\end{equation}
Up to this point everything applies for either side of the brane,
i.e. for $\Lambda=\Lambda_\pm, m=m_\pm, \epsilon=\epsilon_\pm$.

At high energies on the brane, $\rho> \lambda$, where $\lambda$ is
the brane tension, particle interactions produce gravitons that
escape into the bulk. Assuming that all such gravitons are emitted
radially, and that we can neglect their back-reaction on the bulk
geometry, we can use the Vaidya-AdS metric to model the bulk.
(Corrections introduced by allowing for non-radial emission
require numerical computation~\cite{LS}.) The function $\psi$ is
determined in the radiation era ($p=\rho /3$) by kinetic theory
arguments as~\cite{LSR}:
\begin{equation}
\psi =\frac{\kappa_5^{2}}{12}\alpha \rho ^{2}\,, \label{kin}
\end{equation}
where $\alpha $ is a small dimensionless constant. Because of
graviton emission, the brane energy is not conserved,
\begin{equation}
\dot{\rho}+4H\rho =-2\psi \,,  \label{noconti}
\end{equation}
where $H=\dot a/a$ is the Hubble rate.

The radiation leaving the brane is the same on both sides of the
brane, but the different curvature on either side affects the
propagation of the radiation. The parameters $\epsilon _{\pm }$
describe where the radiation is going: $\epsilon=1$ indicates
radiation into the black hole, $\epsilon=-1$ indicates radiation
to infinity. There are four cases: \begin{itemize} \item For
$\epsilon_{\pm }=1$, the brane radiates into black holes on either
side of the brane. \item For $\epsilon_\pm=-1$, the radiation is
escaping to infinity on both sides of the brane. \item For the
remaining two cases, with $\epsilon_+ \epsilon_-=-1$, radiation
falls into a black hole on one side and escapes to infinity on the
other. \end{itemize} Note that in the non-radiating case ($\dot
m=0$), there are also four cases with the above values of
$\epsilon_\pm$, with $\epsilon=1$ denoting the half-bulk that
includes the black hole (or that includes the AdS Cauchy horizons
in the case $m=0$), while $\epsilon=-1$ denotes the half-bulk on
the other side of the brane.

These four cases show important differences in the relation
between the brane energy density and the bulk geometry. The key
equation is Eq.~(77) in~\cite{Decomp}, which in our scenario
becomes~\cite{n1}
\begin{equation} \label{sign}
{\kappa_5^2 a\over 3}(\rho+\lambda)= \epsilon_+\sqrt{\dot{a}^2
+f_+}\,+\, \epsilon_-\sqrt{\dot{a}^2 +f_-}\,.
\end{equation}

The simplest generalization of the $Z_2$-symmetric models
in~\cite{LSR} is the case with $\epsilon_\pm=1$, and in this case,
Eq.~(\ref{sign}) shows that $\rho+\lambda\geq 0$. By contrast, the
$\epsilon_\pm=-1$ models, in which the radiation escapes to
infinity on both sides, have $\rho+\lambda \leq 0$. It is not
possible to have radiation escaping to infinity on both sides
without having negative total energy density on the brane. This
negativity also holds in the non-radiating case and in the
$Z_2$-symmetric case; it is not the asymmetry or the radiation
escaping that causes the problem, but the fact that the bulk
curvature is decreasing as one moves away from the brane. The
models with $\epsilon_+ \epsilon_-=-1$ will in general have
$\rho+\lambda$ changing sign over time. In summary,
\begin{eqnarray}
\epsilon_\pm=1 &\Rightarrow& \rho+\lambda\geq 0\,,\\
\epsilon_\pm=-1 &\Rightarrow& \rho+\lambda\leq 0\,,\\
\epsilon_+ \epsilon_-=-1 &\Rightarrow& \rho+\lambda~\mbox{can
change sign.}
\end{eqnarray}

Equations~(\ref{psi}), (\ref{vdot}) and (\ref{kin}) show how the
mass function evolution at the brane depends on the bulk geometry
on either side of the brane:
\begin{equation}
\dot{m}_{\pm }=\frac{\kappa_5^{4}\alpha }{36}\rho^2 a^{4}\left[ -
H+\epsilon_{\pm }\left( H^{2}+{f_{\pm } \over a^2} \right)
^{1/2}\right]. \label{mdot}
\end{equation}
The average, $\bar m=(m_++m_-)/2$, gives
\begin{eqnarray}
\frac{36}{\kappa_5 ^{4}\alpha a^{3}\rho ^{2}}\dot{\overline{m}}+
\dot a =&&\epsilon_{+}\left( {\dot{a}^{2}+\bar{f}}+\frac{\Delta
f}{
2}\right) ^{1/2} \notag \\
&&{} +\epsilon_{-}\left( {\dot{a}^{2}+\bar{f}}- \frac{\Delta
f}{2}\right) ^{1/2}\!, \label{+}
\end{eqnarray}
where
\begin{eqnarray}
\overline{f} =-\frac{2\overline{m}}{a^{2}}-\frac{\overline{\Lambda
}}{6}a^{2}\,,~ \Delta f =-\frac{2\Delta m}{a^{2}}-\frac{\Delta
{\Lambda }}{6}a^{2}\,.
\end{eqnarray}
The difference, $\Delta m=m_+-m_-$, follows from the square of
Eq.~(\ref{mdot}):
\begin{equation}
\left( \frac{36}{\kappa_5^{4}\alpha a^{4}\rho^{2}}\dot{
\overline{m}}+H\right) \Delta \dot{m}=\frac{\kappa_5 ^{4}\alpha
a^{2}\rho ^{2}}{72 }\Delta f\,.  \label{-}
\end{equation}
The generalized Friedmann equation for an asymmetric brane is
given in~\cite{Decomp}. In our case,
\begin{equation}\label{gf}
H^2=\frac{\kappa_5 ^{4}}{36 }\left( \rho +\lambda \right)
^{2}-{\overline{f}\over a^2}+\frac{9\left( \Delta f\right)
^{2}}{4\kappa_5^4a^{4}(\rho+\lambda)^2}\ .
\end{equation}
This can be employed to eliminate $\dot{a}^{2}+\overline{f}$ from
Eq.~(\ref{+}), giving
\begin{equation}
\frac{36 }{\kappa_5 ^{4}\alpha a^{4}\rho ^{2}}\dot{\overline{m}}+
H =\bar{\epsilon}\,\frac{\kappa_5^{2}\left( \rho +\lambda
\right)}{3} +\Delta \epsilon\, \frac{3\Delta f}
{2\kappa_5^{2}\left( \rho +\lambda \right) a^{2}}. \label{mdot2}
\end{equation}

From now on we will assume that $\rho\geq0$ and $\lambda\geq0$,
and confine our investigation to the $\epsilon_\pm=1$ scenario. In
this case, the projected field equations on the brane~\cite{rev}
show that the effective gravitational constant is
\begin{equation}
\kappa ^{2}\equiv {8\pi \over M_4^2}=\frac{\kappa_5^{4}\lambda
}{6}\,.
\end{equation}
We also fine-tune the cosmological constant on the brane to zero,
\begin{equation}\label{tune}
\kappa^2\lambda+\overline\Lambda=0\,.
\end{equation}

Then the generalized Friedmann equation~(\ref{gf}) and generalized
Raychaudhuri equation~\cite{Decomp} become
\begin{eqnarray}
&&H^2 = \frac{\kappa ^{2}\rho }{3}\left( 1+\frac{\rho }{2\lambda
}\right) +\frac{2\overline{m}}{a^{4}}+ \frac{9\left( \Delta
f\right) ^{2}}{4\kappa_5^4 a^{4}(\rho+
\lambda)^2}\,,  \label{Fr} \\
&&\dot H+H^2 =-\frac{\kappa ^{2}}{3}\rho \left[ 1+\frac{(3+
\alpha)\rho }{2\lambda }\right] -\frac{2\overline{m}}{a^{4}}
\notag
\\
&&~~~~~~~{}-\frac{3\Delta\Lambda\, \Delta f}{2\kappa_5^4 a^2\left(
\rho +\lambda \right)^2 }+\frac{9\left( \rho -3\lambda \right)
\left( \Delta f\right) ^{2}}{ 4\kappa_5^4 a^{4}\left( \rho
+\lambda \right)^3 } \,. \label{Ray}
\end{eqnarray}

In the symmetric case we recover the equations of~\cite{LSR}. The
exact solution is given by~\cite{LMS}:
\begin{eqnarray}
a&=& a_0(t-t_0)^{(2+\alpha)/2(4+\alpha)}
(t-t_1)^{1/(4+\alpha)}\,,\label{s1} \\
\rho &=& {6 \over \kappa^2 A(t-t_0)(t-t_1)}\,,\label{s2}\\ m &=&
{a_0^4 [A(A-8)(t-t_1)+4\ell\{A-2(4+\alpha)\}] \over
8A^2(t-t_0)^{4/(4+\alpha)} (t-t_1)^{\alpha/(4+\alpha)}},\label{s3}
\end{eqnarray}
where the constants are related by
\begin{equation}
t_1=t_0+\ell\,{(4+\alpha) \over A}\,, ~\ell=\sqrt{-\Lambda \over
6}\,.
\end{equation}
In the physically relevant case (i.e., $H\geq0, m\geq0$), we have
$t>t_1>t_0$ and $A\geq 8+2\alpha$.

\section{Solutions}

For the general asymmetric case, we define dimensionless
quantities as follows:
\begin{eqnarray}
\hat{t} &=&\frac{\kappa_5^{2}\lambda }{6}t\,, ~~
\hat{H} =\frac{6}{\kappa_5^{2}\lambda }H \,, \\
\hat{\rho} &=&\frac{\rho }{\lambda } \,,~~ {\hat{m}\over a^4} =
\frac{36}{\kappa_5^{4}\lambda ^{2}}\,{\overline{m}\over a^4} \,,
\\
{q\over a^4} &=&\frac{36}{\kappa_5^{4}\lambda ^{2}}\, { \Delta m
\over a^4}\,,~~ \xi = \frac{|\Delta {\Lambda }|}{\kappa_5
^{4}\lambda ^{2}} \,. \label{varepsilon}
\end{eqnarray}
The first four variables are the same as those used
in~\cite{LSR,LMS}, while $q$ and the parameter $\xi$ encode the
asymmetry.

We constrain $\xi$ by the requirement that ${\Lambda}_{\pm}<0$.
This means that $\Delta \Lambda < \overline{\Lambda}$, so that the
cosmological constants on both sides of the brane are negative.
Then from Eqs.~(\ref{tune}) and (\ref{varepsilon}), we find that
\begin{equation}\label{xi}
0\leq \xi<{1 \over 6}\,.
\end{equation}

Using these variables, the dynamics on the brane in the case
$\epsilon_\pm=1$ are governed by the system of first order linear
differential equations, Eqs.~(\ref{noconti}), (\ref{-}), (\ref
{mdot2}) and (\ref{Ray}). In normalized form this system becomes
\begin{eqnarray}
\hat{\rho}'&=& -4\hat{H}\hat{\rho} -\alpha\hat{\rho}^2 , \\
{\hat{m}'} &=& {\alpha a^4 \hat{\rho}^2} \left(\hat{\rho}+1-\hat
H\right)
\,,\label{dmhat} \\
q'&=& -\alpha {\hat{\rho}^2(q + 3\xi a^4) \over (\hat{\rho}+1)}
\,,  \label{deltamhat} \\ \hat{H}'&=& -\hat{H}^2 -
2\hat{\rho}\left[1+\frac{(3+\alpha)}{2}\hat{\rho} \right]
- \frac{2\hat{m}}{a^4}  \notag \\
&&~{}+\frac{3{\xi}(q+3{\xi} a^4)}{a^4(\hat\rho+1)^2} +
\frac{(\hat{\rho}-3)(q+3{\xi }a^4)^2} {4a^8(\hat{\rho}+1)^3}.
\label{Raydri2}
\end{eqnarray}
The algebraic constraint from the generalized Friedmann equation,
\begin{equation}
\hat{H}^2 =\hat{\rho}\left( 2+\hat{\rho}\right)
+\frac{2\hat{m}}{a^{4}}+ \frac{\left( q+3{\xi}a^4\right)
^{2}}{4a^{4}(\hat{\rho}+1)^2}, \label{Fr2}
\end{equation}
is used to monitor numerical errors. Results from a numerical
integration of this system with values of $\xi$ in the range of
Eq.~(\ref{xi}) and a variety of initial conditions for $q$ are
shown in Figs.~1--5.

\begin{figure}[ht!]
\begin{center}
\includegraphics[height=3in,width=3in,angle=-90]{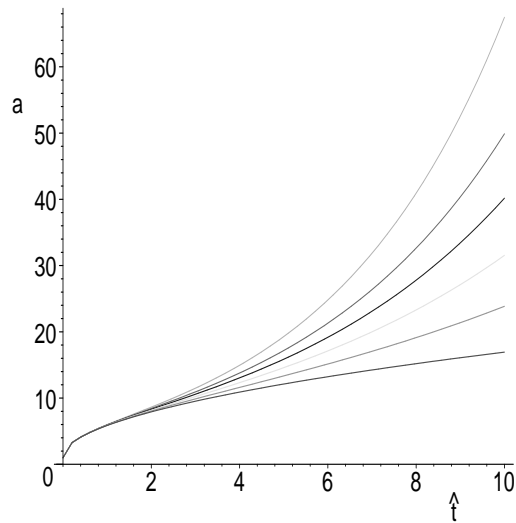}
\end{center}
\caption{The scale factor $a(\hat t)$ with $q(0)=0$. The curves
from bottom to top have $\protect\xi$ increasing from $0$ to
${1\over 6}^-$. } \label{epsilonagraph}
\end{figure}

The first three graphs show the effect of increasing $\xi$ while
keeping $q(0)$ fixed. The effect on the scale factor is to
introduce acceleration at late time. This effect is independent of
the Vaidya radiation and is also found in the non-radiating
case~\cite{lamasym2,both1}. (See~\cite{lamasym2,lamasym1,perk} for
asymmetry in bulk cosmological constants, \cite{bhasym} for
asymmetry in (constant) bulk black hole masses,
and~\cite{both1,both2} for both types of asymmetry). The effects
of the radiating brane occur predominantly at early times.

\begin{figure}[ht!]
\begin{center}
\includegraphics[height=3in,width=3in,angle=-90]{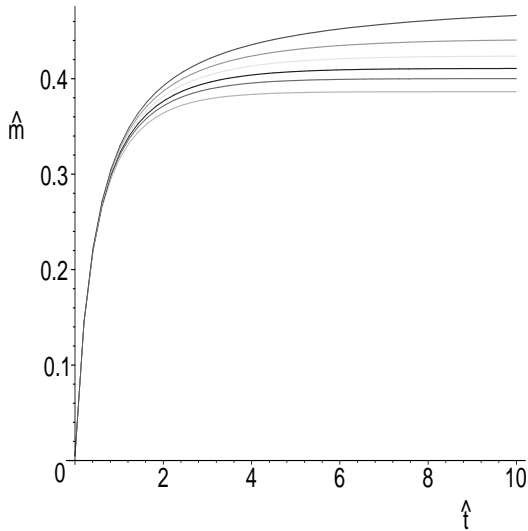}
\end{center}
\caption{The average black hole mass function $\hat{m}(\hat t)$
with $q(0)=0$. The curves from top to bottom have $\protect\xi$
increasing from $0$ to ${1\over 6}^-$. } \label{epsilonmgraph}
\end{figure}

There is also a marked effect on the average mass function
$\hat{m}$. The increase of $\hat H$ when $\xi$ is increased in
Eq.~(\ref{Fr2}) leads to a decrease in $\hat{m}'$ by
Eq.~(\ref{dmhat}), which results in a suppression of the
asymptotic value of $\hat m$. Since the asymptotic value
determines the dark radiation at nucleosnthesis (where
$\rho\ll\lambda$), this suppression of $\hat{m}$ means that
nucleosynthesis limits~\cite{LSR} can be more easily satisfied
when there is asymmetry.

\begin{figure}[ht!]
\begin{center}
\includegraphics[height=3in,width=3in,angle=-90]{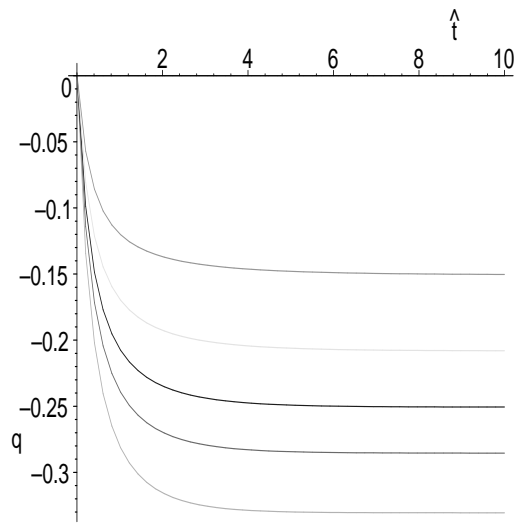}
\end{center}
\caption{The difference in black hole masses, $q(\hat t)$, with
parameters as in Fig.~\ref{epsilonmgraph}} \label{epsilondmgraph}
\end{figure}

Figure~\ref{epsilondmgraph} shows that even if there is no initial
difference in the black hole mass functions on either side of the
brane, and given that the radiation from the brane is symmetric, a
difference in the mass functions will still develop during the
evolution, because of the different curvatures on either side of
the brane.

\begin{figure}[ht]
\begin{center}
\includegraphics[height=3in,width=3in,angle=-90]{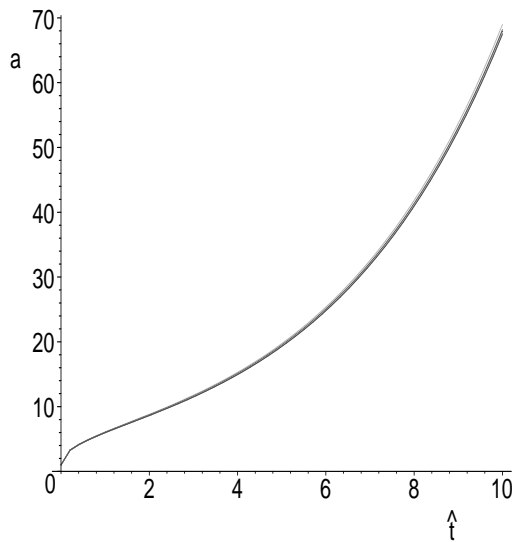}
\end{center}
\caption{The scale factor $a(\hat t)$ with $\protect\xi={1\over
6}^-$ and $q(0)=0$, $10$, $20$, $30$, $50$ and $100$, from bottom
to top curve. Variation in $q(0)$ has little effect on the
evolution of $a$.} \label{dma36graph}
\end{figure}

\begin{figure}[ht]
\begin{center}
\includegraphics[height=3in,width=3in,angle=-90]{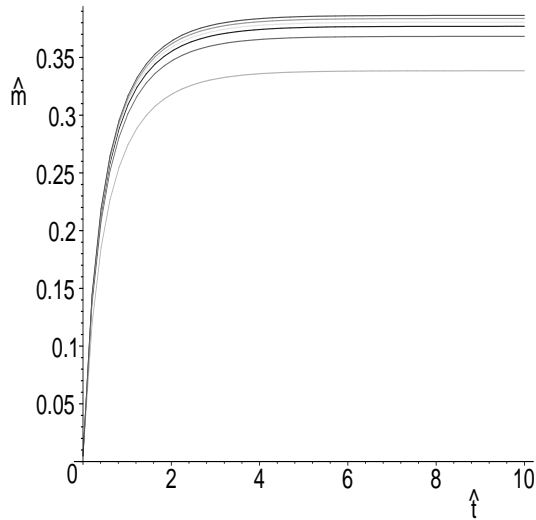}
\end{center}
\caption{The average black hole mass function $\hat{m}(\hat t)$
with $\protect\xi=0 $ and (curves from top to bottom) $q(0)=0$,
$10$, $20$, $30$ , $50$ and $100$.} \label{dmm36graph}
\end{figure}

Figures~\ref{dma36graph} and \ref{dmm36graph} demonstrate the
effect of varying $q(0)$ for a fixed $\xi$. Although there is
little discernible difference in the scale factor for the
different values of $q(0)$, the increase in $\hat{H}$ for larger
$q(0)$ is still enough to suppress the final value of $\hat{m}$.
(The shape of the $q$ curve is the same as in
Fig.~\ref{epsilondmgraph} for all the initial values.)

If we assume that the asymmetry is small, i.e. $q/a^4$ and $\xi$
are small, then we can perturb away from the symmetric solution,
Eqs.~(\ref{s1})--(\ref{s3}), and evaluate the early- and late-time
solutions. At early times,
\begin{equation}
{q\over a^4} \approx \frac{3\alpha }{4+\alpha} \left(1 + {t_2\over
t} \right)\,,
\end{equation}
where $t_2$ is a constant of integration which must be of a
suitable size to keep the whole expression within our smallness
assumptions. This type of decay at early times is confirmed in
Fig.~\ref{epsilondmgraph}. At late times,
\begin{equation}
{q} \approx  Be^{-C/t^2}\,,
\end{equation}
where $B,C$ are constants. The asymptotic value, $q\to B$, is
evident in Fig.~\ref{epsilondmgraph}.

\section{Conclusion}

We have derived the dynamical equations for a radiating brane in a
Vaidya-AdS bulk, without assuming $Z_2$ symmetry. The radiation
from the brane propagates in different curvature regions on either
side, given the difference in bulk cosmological constants and
initial black hole masses on either side. There are four cases,
depending on whether the radiation falls into a black hole or
escapes to infinity on either side of the brane.

We performed numerical integrations of the equations in the case
where radiation feeds black holes on both sides of the brane. One
of the key features that emerges is the suppression of the
late-time dark radiation (the asymptotic value of the average mass
parameter) due to asymmetry.

These solutions correspond to the case $\epsilon_\pm=1$ in
Eq.~(\ref{sign}), or $\overline\epsilon=+1, \Delta\epsilon=0$ in
Eq.~(\ref{mdot2}). The other three cases have radiation escaping
to infinity on at least one side of the brane, but in these cases,
the total brane energy density $\rho+\lambda$ is negative or can
change sign. Further investigation is needed to determine whether
any of these cases admit physically reasonable models.

\[ \]{\bf Acknowledgments:}

L\'{A}G was supported by OTKA grants Nos. T046939, TS044665 and by
a PPARC visitor grant, EL by an EPSRC DTA grant, and RM by PPARC.


\begin{thebibliography}{9}
\bibitem{n1}
This generalizes Eq.~(21) in~\cite{lamasym1}, which covers
non-radiating $\epsilon_+ \epsilon_-=-1$ models.
\end{thebibliography}
\end{document}